\def\umass{1}
\def\upc{2}
\def\ieec{3}
\def\ucsb{4}
\def\exeter{5}

\documentclass[iop]{emulateapj}
\usepackage{natbib}
\usepackage{graphicx}
\usepackage[space]{grffile}
\usepackage{latexsym}
\usepackage{amsfonts,amsmath,amssymb}
\usepackage{url}
\usepackage[utf8]{inputenc}
\usepackage{fancyref}
\usepackage{hyperref}
\hypersetup{colorlinks=false,pdfborder={0 0 0},}

\begin{document}

\title{Spiral Disk Instability Can Drive Thermonuclear Explosions in Binary White Dwarf Mergers}

%% Notice placement of commas and superscripts and use of &
%% in the author list

\author{Rahul Kashyap$^{1}$, Robert Fisher$^{1}$, Enrique Garc\'{i}a-Berro$^{2,3}$, Gabriela Aznar-Sigu\'{a}n$^{2,3}$, Suoqing Ji$^{4}$, Pablo Lor\'{e}n-Aguilar$^{5}$}

%\maketitle

\altaffiltext {\umass} {Department of Physics, University of Massachusetts Dartmouth, 285 Old Westport Road, North Dartmouth, MA. 02740, USA}
\altaffiltext {\upc} {Departament de F\'{i}sica Aplicada, Universitat Polit\`{e}cnica de Catalunya, c/Esteve Terrades, 5, E-08860 Castelldefels, Spain}
\altaffiltext {\ieec} {Institut d'Estudis Espacials de Catalunya, Ed. Nexus-201, c/Gran Capit\`a 2-4, E-08034 Barcelona, Spain}
\altaffiltext {\ucsb}  {Department of Physics, Broida Hall, University of California Santa Barbara, Santa Barbara, CA. 93106-9530, USA}
\altaffiltext {\exeter} {School of Physics, University of Exeter, Stocker Road, Exeter EX4 4QL, UK}

\begin {abstract}
Thermonuclear, or Type Ia supernovae (SNe Ia), originate from the explosion of carbon--oxygen white dwarfs, and serve as standardizable cosmological candles. However, despite their importance, the nature of the progenitor systems that give rise to SNe Ia has not been hitherto elucidated. Observational evidence favors the double-degenerate channel, in which merging white dwarf binaries lead to SNe Ia. Furthermore, significant discrepancies exist between observations and theory, and to date, there has been no self-consistent merger model that yields a SNe Ia. Here we show that a spiral mode instability in the accretion disk formed during a binary white dwarf merger leads to a detonation on a dynamical timescale. This mechanism sheds light on how white dwarf mergers may frequently yield SNe Ia. 
\end {abstract}


\section {Introduction}
It has long been suggested that a white dwarf in a binary system -- with either another white dwarf, through the so-called double-degenerate channel \citep {Iben_Tutukov_1984}, or a main-sequence or red giant companion through the single-degenerate channel \citep {Whelan_Iben_1973} -- could give rise to a Type Ia supernova (SN Ia) event. Yet, despite the recent breakthrough detections of relatively nearby SNe Ia, including 2011fe \citep{Nugent_2011} and 2014J \citep{Zheng_2014}, no normal SNe Ia stellar progenitor has ever been directly identified.  Overall, the challenging, unsolved stellar progenitor problem for SNe Ia stands in marked contrast to the case of core-collapse supernovae, whose stellar progenitors are highly luminous, massive stars.

In light of the challenges to the single-degenerate channel of SNe Ia, increasing attention is being paid to the double-degenerate channel. In contrast to the single-degenerate channel, the double-degenerate channel offers natural solutions to the stellar progenitor problem, as well as to other problems: the delay--time distribution \citep {maozbadenes_2010}, the observed absence of dense circumstellar gas surrounding the progenitor \citep {badenes_etal_2007}, and the absence of the surviving binary member star in SNe Ia remants \citep {kerzendorf_etal_2014}. 

However, the double-degenerate channel in turn presents its own set of challenges. From an observational and a theoretical perspective, one of the most outstanding challenges is to understand the conditions under which a merging white dwarf binary system yields a detonation of the underlying carbon--oxygen white dwarf, and hence a SNe Ia. On this issue, however, there currently exists a significant discrepancy between observations and theory. On the one hand, if double-degenerate mergers must account for the observed SNe Ia rate, observations appear to require that the majority of mergers, including sub-Chandrasekhar mass systems, yield SNe Ia \citep {badenesmaoz12}. In contrast, the most extensive numerical studies completed to date suggest that essentially no carbon-oxygen white dwarf binary systems self-consistently achieve detonation conditions during the merger phase of evolution \citep {Zhu_2013, Dan_etal_2014}. Simulations of the detonation phase of the so-called ``violent mergers'' of the most massive of these systems rely instead upon artificially igniting a detonation front  \citep {Pakmor_2010, Pakmor_2012, Raskin_2014, Moll_2014}. A proposed model involving head-on collisions of white dwarfs \citep {Hawley_etal_2012,Raskin_2009,Kushnir_2013,Aznar_2013,Aznar_2014} demonstrates robust detonations. However, it remains unclear whether head-on collisions, even in triple systems, match observed SNe Ia rates and $^{56}$Ni yields \citep {Scalzo_2014}.

For all prompt merger scenarios, it is crucial to recognize that the $^{56}$Ni nucleosynthetic yield, and hence the peak luminosity, is closely correlated with the central density, or equivalently, the mass of the white dwarf primary \citep {Raskin_2014}.  Consequently, in the context of double-degenerate mergers, a deeper understanding of the most fundamental properties of SNe Ia, including both the absolute value and relatively small variance  of their intrinsic luminosity, hinges crucially upon an explication of their detonation mechanism.

\section{Numerical Simulations}

We conducted a set of smoothed particle hydrodynamics (SPH) simulations of merging carbon--oxygen white dwarfs, with equal abundances of carbon and oxygen, and with masses of $1.0 M_{\odot}$ and $1.1 M_{\odot}$. For such simulations we used the SPH code employed in previous studies of this kind \citep{Loren_Aguilar_2010}, with a resolution of $4\times 10^5$ SPH particles. The overall time evolution of the merger process found in these simulations and the general appearance of the remnants are very similar to those found in other previous studies. 

At approximately the instant at which the maximum temperature is reached during the merger, the SPH conditions are remapped onto an Eulerian adaptive mesh, using the FLASH4 code framework \citep{Fryxell_2000,arch2009}. For the model presented here, this remapping occurs 1.5 initial orbital periods, or 40 s, subsequent to tidal disruption. Throughout this paper, we refer to the time of remapping as $t = 0$.  

FLASH solves the Euler equations of hydrodynamics using an unsplit hydrodynamics solver \citep{Lee_2009}, and closed with an equation of state that includes contributions from blackbody radiation, ions, and electrons of an arbitrary degree of degeneracy \citep{Timmes_2000}. Nuclear burning is incorporated through the use of a simplified 13 species alpha-chain network, which includes the effect of neutrino cooling \citep{Timmes_1999}. The Poisson equation for self-gravity is solved both with a recently improved multipole solver \citep{Couch_2013}, which treats gravitational potential at cell faces and centers the expansion at a square-density-weighted mean location, and also for comparison purposes, with an oct-tree direct multigrid method \citep{Ricker_2008}.

\begin{figure}
\begin{center}
\includegraphics[width=0.7\columnwidth]{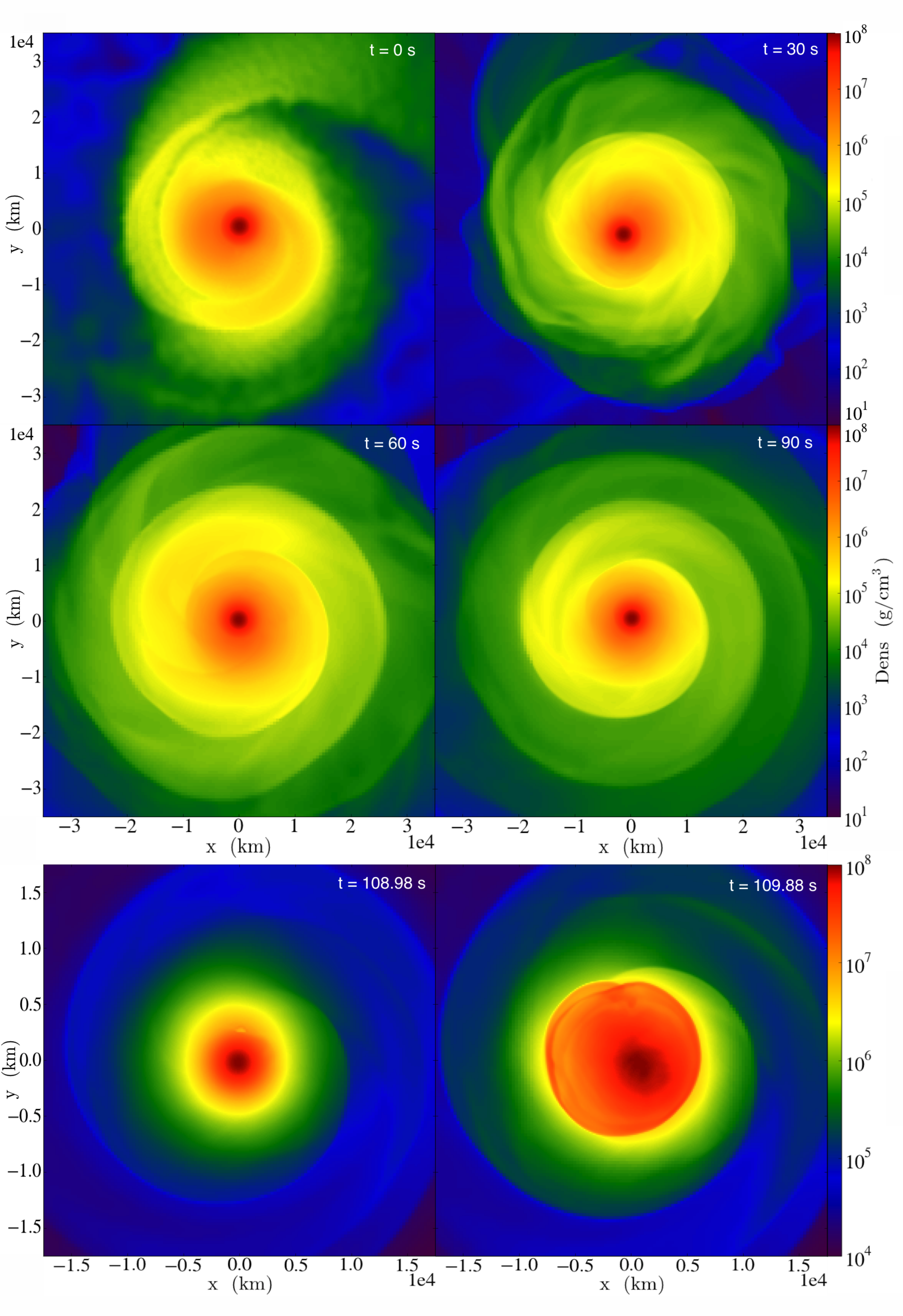}
\caption{ Time evolution of the binary white dwarf system subsequent to merger. This figure presents the time evolution of the density in the mid-plane of the disk, shown on a logarithmic scale. The bottom two frames, taken at $t = 108.98$ and $109.88$ s, are just prior to and just subsequent to the onset of detonation, respectively. The bottom two frames are zoomed-in by a factor of two in the $x$ and $y$ directions. }
\label{fig:time_evol}
\end{center}
\end{figure}

Like others \citep {Pakmor_2010, Pakmor_2012, Raskin_2014, Moll_2014}, we do not find that the initial merger conditions produce a self-consistent detonation at the onset of the three-dimensional FLASH simulation. However, unlike prior work, we choose not to initiate a detonation artificially and instead follow the subsequent evolution of the system forward in time. Figure ~\ref{fig:time_evol} displays six snapshots of the overall evolution of the merger remnant. By $t = 30$ s, or roughly one initial orbital period, a pronounced $m = 1$ spiral mode develops, which persists for several dynamical times.  At $t = 108.98$ s, a strong detonation driven by the spiral mode occurs self-consistently, giving rise to a supernova outburst.

\section{Gravitational Instability and Spiral modes}
During the final inspiral of the binary white dwarf system, as the less massive secondary white dwarf approaches the more massive primary, it is tidally disrupted, producing an accretion disk, with little mass ejected from the system. Conditions for thermonuclear detonation in degenerate carbon require hot temperatures $T \sim 2 \times 10^9$ K at densities $\rho \sim 10^7$ g/cm$^3$ \citep {Seitenzahl_2009}. In a sufficiently massive system, the primary has the requisite density for detonation, and the disk has the necessary temperature. However, for unequal mass mergers, the tidal disruption of the secondary leaves the primary largely intact and unheated. Consequently, a key challenge faced by double-degenerate models is the need for some mechanism to raise the temperature of the cold primary ($T \sim 10^7$ K) in order to produce a SNe Ia. If the mass accretion rate onto the central white dwarf is sufficiently rapid, it may mix hot, tidally disrupted disk material with the higher density white dwarf primary material, leading to unstable carbon burning and a subsequent detonation, as previous studies of single-degenerate models have revealed \citep {Jordan_2012}.

Fundamental physical considerations determine the basic physical state and evolution of the gas in the accretion disk, as a function of the primary mass $M_1$, the secondary mass $M_2$, as well as the mass ratio $q = M_2 / M_1$. In particular, the secondary mass is disrupted as its orbit approaches the tidal radius $R_{\rm tid} = q^{-1/3} R_2$, defined as the orbital separation distance at which the mean density of the primary, averaged over the entire spherical volume of the orbit, is equal to that of the secondary star, with an initial undeformed radius of $R_2$. 

We now show that the inner disk of a white dwarf merger is marginally gravitationally stable and subject to the development of non-axisymmetric spiral modes, based upon the physics of tidal disruption of the secondary. We assume that the secondary is completely disrupted and extends radially from the tidal radius $R_{\rm tid}$ to large radii, while the primary remains intact. These assumptions are appropriate for mergers with unequal masses, which we verify {\it a posteriori} from numerical simulation data. 
  
The sound speed $c_s$ within the disk is then directly determined from the thermal energy,  $c_s (r) = (f \gamma GM_1 /  r)^{1/2}$ in the non-degenerate ideal gas conditions prevalent in the disk,  where $\gamma \simeq 5/3$ is the ratio of specific heats, and $f \sim 0.2$ is the fraction of the maximum virial temperature, as determined from numerical simulations \citep{Zhu_2013}.   
 In the inner disk, the rotational profile is nearly Keplerian, and the epicyclic frequency is $\kappa (r) \simeq \Omega (r) \simeq (G M_1 / r^3)^{1/2}$.  In addition, the surface gravity of the disk is determined by the surface density $\Sigma (r)$ of the disk, which falls off as a power-law in radius, $r^{-\eta}$, with a relatively steep scaling of $\eta \simeq 2.5 - 3.5$, also determined from numerical simulations \citep{fryeretal_2010}.  In the following, we presume a typical surface density scaling $\Sigma = (M_2  /  2 \pi R_{\rm tid}^2) ({R_{\rm tid} / r})^{3}$.  

These fundamental physical properties of the disk can then be combined to form the Toomre parameter $Q = c_s \kappa/ \pi G \Sigma$, which quantifies the degree to which the disk is subject to gravitational instability. Here, $\kappa$ is the epicyclic frequency, with $\kappa \sim \Omega$ for a near-Keplerian disk. In particular, a classical analytic result yields that if $Q < 1$, disks are subject to axisymmetric instability. 
 
Furthermore, models of massive accretion disks have demonstrated the existence of an eccentric gravitational instability even in disks that are otherwise gravitationally stable for disk masses exceeding a third that of the central stellar mass, provided that $Q < 3$ at corotation \citep {Adams_1989,Shu_1990}. The eccentric gravitational instability gives rise to a characteristic $m = 1$ spiral mode, which was originally studied in the context of massive star-forming accretion disks and has subsequently been examined in a variety of other contexts \citep {Hopkins_2010}.

On the basis of these basic physical considerations, $Q (r) \sim q^{-1} (r / R_{\rm tid} )$ for binary white dwarf mergers, indicating marginal gravitational stability for a typical merger system. Furthermore, we note that this scaling of $Q$ values is, remarkably, independent of the total system mass. Consequently, we infer that many binary dwarf mergers, including both sub-Chandraskehar and super-Chandrasekhar mass systems, with sufficiently large mass ratio $q$ will generally be subject to spiral mode instabilities, including the eccentric gravitational instability.

\begin{figure}
\begin{center}
\includegraphics[width=0.9\columnwidth]{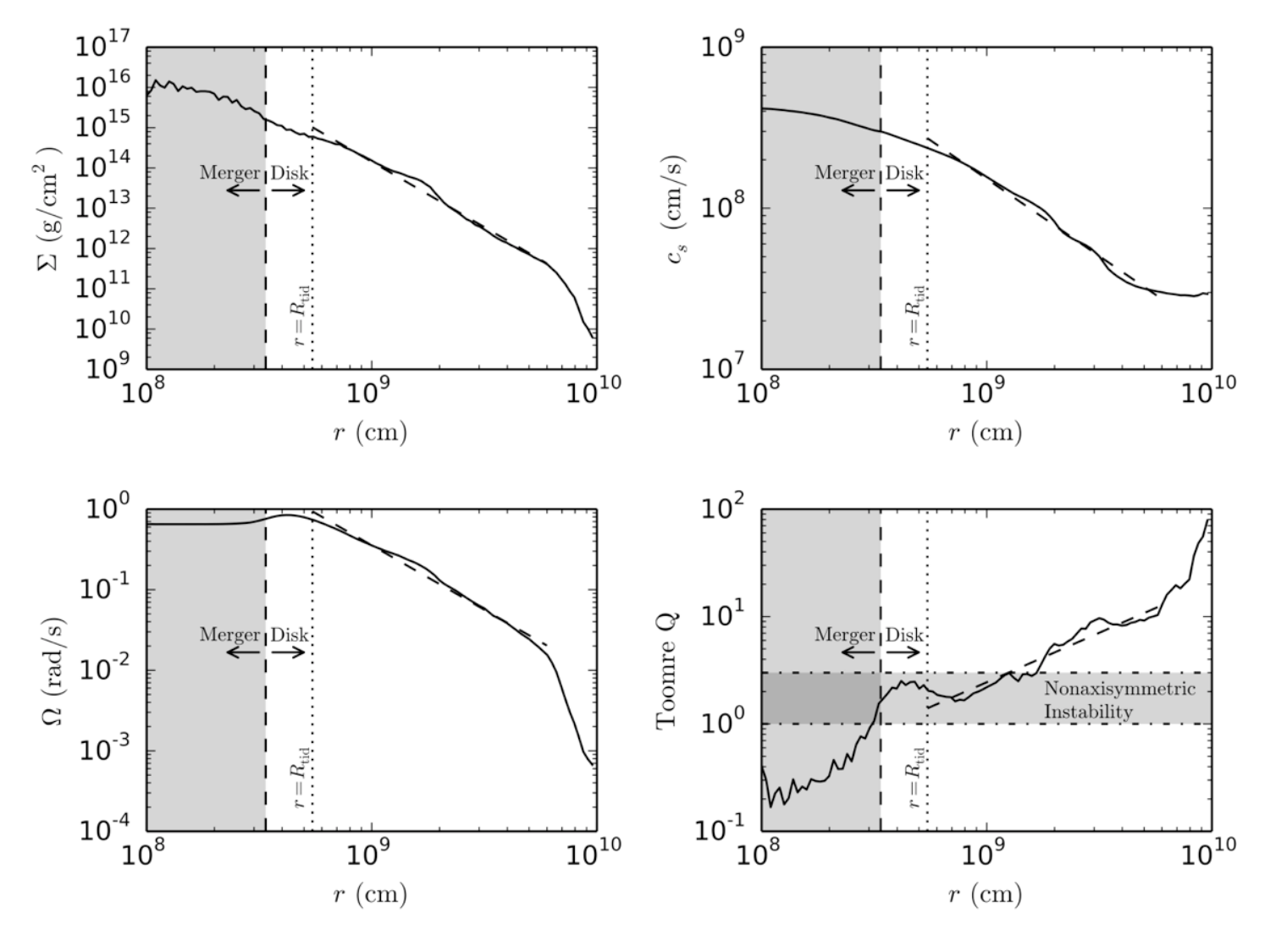}
\caption{Properties of the accretion disk. Clockwise, from upper-left, this plot presents surface density $\Sigma$, sound speed $c_s$, angular velocity $\Omega$, and Toomre $Q$ parameter versus cylindrical radius $r$ as measured from the peak central density of the white dwarf merger, all on logarithmic--logarithmic axes, for the initial time $t = 0$ at the beginning of the simulations. The vertical dashed line indicates the radial location enclosing $M_1 = 1.1 M_{\odot}$, which demarcates the primary white dwarf on the left from the disk on the right. For comparison, we also indicate with a dashed line  the location of $r = R_{\rm tid}$. The power law fits have indices of -3.25 for $\Sigma$, -0.96 for $c_s$, -1.60 for $\Omega$, and 0.92 for Toomre $Q$. The fits are in very good agreement with the simple scalings presented in the text.}
\label{fig:disk_properties}
\end{center}
\end{figure}

To elucidate the properties of the disk, we analyze the properties of the white dwarf merger system at $t = 0$ in figure~\ref{fig:disk_properties}.  The values and the power law scalings of the surface density $\Sigma (r)$, sound speed $c_s (r)$, angular velocity $\Omega (r)$, and   $Q (r)$ as determined from the full numerical simulation of the merger process are in very good agreement with our analytic estimates. In particular, $Q \simeq 1$ in the inner accretion disk, confirming our prediction that the inner disk is marginally gravitationally stable.

\begin{figure}
\begin{center}
\includegraphics[width=0.9\columnwidth]{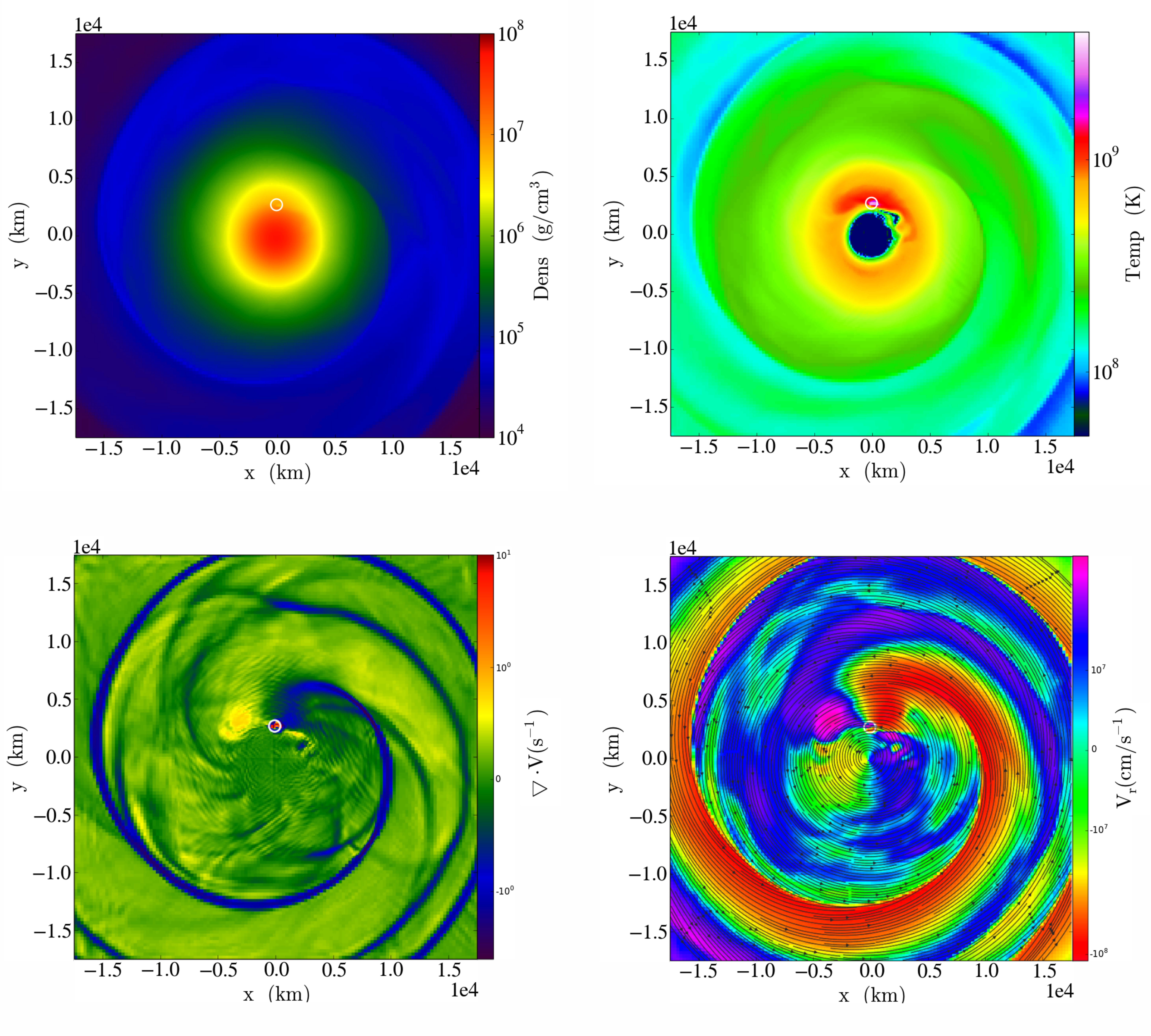}
\caption{Detailed view of the detonation.  All four panels depict the first instant of the detonation tracked in the simulation, at $t = 108.98$ s, corresponding to the bottom left panel of figure~\ref{fig:time_evol}. The upper-left panel shows the log($\rho$), and shows the location of the detonation, near $\sim 10^7$ g/cm$^3$, as indicated by the white circle. The upper-right panel shows the log($T$) where the peak temperature has risen to $\sim 3 \times 10^9$ K at the location of detonation. The bottom-left panel shows the divergence of the velocity field; positive values represent diverging flows and negative values indicate converging flows. The strongly negative values coincide with the spiral density waves of the first panel, and  demonstrate the existence of shocks at the location of the spirals.  The bottom-right panel shows the velocity streamlines in the disk, shaded by the radial velocity. The spiral structure directly gives rise to a funnel accretion inflow onto to the white dwarf primary, which mixes hot disk material into the merger and gives rise to the detonation.}
\label{fig:det_time}
\end{center}
\end{figure}

We further characterize the detonation mechanism in figure~\ref{fig:det_time}. This figure analyzes the structure of the merger at the instant of detonation.  As previous studies \citep {Blaes_1988} have demonstrated, the spiral instability provides a mechanism that drives an inflow of hot material from the inner disk inward. The connection of the spiral to the funnel accretion flow is made clear in the bottom two panels of figure~\ref{fig:det_time}. The hot matter from the tidally disrupted secondary is accreted inward through a funnel accretion flow generated by the spiral instability.

\section{Detonation and Numerical Tests}

The detonation is initiated in the degenerate material of the primary,  where the density $\rho = 6.7 \times 10^6$ g/cm$^3$, and the temperature has been raised to $T  = 3.2 \times 10^9$ K just prior to runaway through the admixture of degenerate white dwarf and hot disk material. Such a detonation arises due to the high virial temperature of the disk. Using a Salpeter IMF and the empirical initial-to-final white dwarf mass relation of \cite{Kalirai_2008}, we estimate that the fraction of WD binaries with a total mass greater than 2.1 $M_{\odot}$ to be 2\%. A much larger fraction of binaries will be susceptible to unstable helium burning (R. Kashyap et al. 2015, in preparation).

The simulation undergoes a nuclear runaway, and the overpressure from the resulting nuclear energy injection initiates the detonation front. The minimum  detonation length scale is 7 km, and is much smaller still for the carbon-burning zone, of cm order at these densities and temperatures \citep {Seitenzahl_etal_2009}. Both are clearly unresolved in all existing 3D simulations. However, we emphasize that the initiation is driven self-consistently by the spiral mode here without the need for an artificial detonation of any sort.  Furthermore, an extensive set of verification tests demonstrates that the key properties of the detonation are all robustly determined above a critical numerical resolution required to resolve spiral structure. A total of $\sim 0.64 - 0.72 M_{\odot}$ of $^{56}$Ni  is synthesized during the detonation, corresponding to a normal-brightness SN Ia event.

We have carried out extensive verification tests to confirm the reliability and reproducibility of our simulations. In particular, we have carried out a convergence test, using higher and lower resolutions than our standard model. Additionally, because the use of the multipole gravity solver could be called into question in a strongly non-spherical flow, we repeated our simulations using an entirely independent multigrid gravity solver. The runs are summarized in table \ref{runtable}, where they are listed in order of decreasing resolution. Additionally, runs designated ``mp'' utilize the multipole solver, whereas ``mg'' runs employ multigrid. The domain size in all models is held fixed.

\begin{deluxetable}{cccccc}
\tablecolumns{8}
\tablewidth{0pc}
\tablecaption{Runs in This Paper. Maximum finest resolution for each model is quoted along with the time of detonation (if any). Models that do not detonate list a dash (---) in place of a time of detonation, $M (^{56}$Ni), and  $\max (E_{\rm kin})$.}
\tablenum{1}
\tablehead{\colhead{Run} & \colhead{Resolution} & \colhead{$c_{\rm burn}$} & \colhead{$t_{\rm det}$} & \colhead{$M (^{56}$Ni)} & \colhead{$\max (E_{\rm kin})$} \\ 
\colhead{} & \colhead{(km)} & \colhead{} & \colhead{(s)} & \colhead{($M_{\odot}$)} & \colhead{(erg)} } 
%\begin{tabular}{|c|c|c|c|c|c|}
\startdata

1-mp  &  68.3 km  & 1   & 89 s    & 0.72 $M_{\odot}$ & $1.10 \times 10^{51}$ \\
2-mp1 &  136 km   & 1   & 112 s   & 0.70 $M_{\odot}$ & $1.08 \times 10^{51}$ \\
2-mp2 &  136 km   & 0.5 & 112 s   & 0.66 $M_{\odot}$ & $1.08 \times 10^{51}$ \\
2-mp3 &  136 km   & 0.3 & 112 s   & 0.64 $M_{\odot}$ & $1.08 \times 10^{51}$ \\
3-mp  &  273 km   & 1   & ---     &  ---             & --- \\ 
4-mp  &  546 km   & 1   & ---     & ---              & --- \\ 
1-mg  &  68.3 km  & 1   & 93.1 s  & 0.72 $M_{\odot}$ & $1.06 \times 10^{51}$  \\
2-mg  &  136 km   &  1  & 107 s   & 0.70 $M_{\odot}$ & $1.08 \times 10^{51}$  \\
3-mg  &  273 km   & 1   & ---     & ---              & --- \\

\enddata
\label{runtable}
\end{deluxetable}%

At sufficiently high resolution, all models using multigrid and multipole produce nearly identical spiral structures and detonations. More importantly, these models demonstrate that the results are entirely insensitive to the choice of gravity solver at sufficiently high resolution. Further, these models demonstrate convergence at the level of $\sim 3$\% in the quantity of $^{56}$Ni produced. Lower resolution models (3-mp, 3-mg, 4-mp) fail to detonate. Analysis of these models reveals that the spiral structure is much less pronounced in runs 3-mp and 3-mg and entirely absent from 4-mp. These findings are consistent with the interpretation that the detonation in these models is driven by the spiral structure generated in the accretion disk.

Additionally, we have conducted a convergence study in the time domain by reducing the chosen timestep for a fixed spatial resolution.  In particular, the timestep $\Delta t^n$ in the $n$th timestep is chosen to be the more restrictive of the CFL timestep $\Delta t^n_{\rm hydro}$ and the burn timestep $\Delta t^n_{\rm burn}$, e.g. $\Delta t^n = \rm {min} (\Delta t^n_{\rm hydro}, \Delta t^n_{\rm burn}  )$. The burn timestep is in turn restricted to a value that prohibits the internal energy $e_{\rm int}$ of a given cell from changing by a fraction $c_{\rm burn}$ in a given timestep. That is, $\Delta t^n_{\rm burn} =  \Delta t^{n - 1} c_{\rm burn}  (e_{\rm int} / \dot {e}_{\rm nuc} \Delta t^{n -1} )   $, where $\dot {e}_{\rm nuc}$ is the specific rate of change of internal energy due to nuclear processes.

Previous studies of zero-impact parameter colliding white dwarfs \citep {Hawley_etal_2012} demonstrated a high degree of sensitivity to the burning time step criterion, through the parameter $c_{\rm burn}$, with $^{56}$Ni yields varying by roughly 30\%. These calculations suggested that the amount of $^{56}$Ni generated through collisions is sensitive to the resolution of the simulation in the time domain. To check this dependence, we computed three sets of models for our standard case, varying  $c_{\rm burn}$ from 1 (effectively using the CFL condition only) to 0.5 and 0.3. We find a much stronger degree of sensitivity to the timestep selection than the spatial resolution, with the overall $^{56}$Ni yield varying by slightly less than 10\% over the range of models considered, with more highly-refined time domain calculations producing less $^{56}$Ni. However, overall we find a smaller degree of sensitivity to the choice of the burning timestep than \citet {Hawley_etal_2012}, and indeed there is a trend towards convergence evident in the more highly refined time domain runs. The use of the burning-adjusted timestep criterion dramatically reduces the timestep relative to a calculation adopting the CFL condition only. In our simulations, the minimum burning-adjusted timesteps for $c_{\rm burn} =$ 0.5 and 0.3 dropped to factors of $10^{-1}$ and $10^{-3}$, relative to the CFL timestep. Further reductions in $c_{\rm burn}$ to 0.2 dropped the timestep to a factor of $10^{-5}$ relative to the CFL timestep, and proved to be computationally intractable.

We note that the structure of the nuclear network involves a stiff reaction solver which evolves the nuclear reactions at every $\Delta t$. However, the hydrodynamics is (by definition) not subcycled for an unsplit explicit solver. Consequently, the choice of the timestep directly determines not so much the accuracy of the burning as  the accuracy of hydrodynamical coupling to the nuclear burning.  In this capacity, it is not surprising that the head-on collision of white dwarfs, involving very strong shock--shock interactions on very small length scales, demonstrates a more sensitive coupling of the hydrodynamics to nuclear burning.
% than the relatively more mild hydrodynamical processes of a merger. In this paper we have followed the mid-term evolution of the final product of the merger of a double white dwarf system.

\section{Conclusion}
We have demonstrated that the innermost region of the accretion disk formed in the aftermath of a sufficiently massive white dwarf binary merger with a large mass ratio is accreted on the primary white dwarf due to spiral instabilities. The ensuing hot disk material accreted on to the central remnant drives a powerful thermonuclear outburst, leading to a SN Ia explosion, without the need for artificial ignition. The spiral mechanism operates on a dynamical timescale, about two to three orders of magnitude more rapidly than an alternative mechanism driven by the magnetorotational instability \citep {Ji_2013}.
Our results demonstrate the robustness of the double-degenerate scenario over a wider range of mass ratios than previously thought possible  and shed light on the possible scenarios producing SN Ia, a topic of the major interest because of their application to cosmology.

{\bf Acknowledgements}  We thank James Guillochon, Lars Bildsten, Matthew Wise, and Gunnar Martin Lellep for useful discussions, and Matthias Aegenheyster for his contributions to the FLASH analysis codes. EGB acknowledges support from MCINN grant AYA2011--23102, and from the European Union FEDER fund. The software used in this work was in part developed by the DOE NNSA-ASC OASCR Flash Center at the University of Chicago. This work used the Extreme Science and Engineering Discovery Environment (XSEDE), which is supported by National Science Foundation grant number ACI-1053575. Simulations at UMass Dartmouth were performed on a computer cluster supported by NSF grant CNS-0959382 and AFOSR DURIP grant FA9550-10-1-0354. This research has made use of NASA's Astrophysics Data System and the yt astrophysics analysis software suite \cite{Turk_2011}. RTF is grateful to have had the opportunity to complete this paper during a visit to the Kavli Institute for Theoretical Physics, which is supported in part by the National Science Foundation under Grant No. NSF PHY11-25915.

\end{document}